\documentclass[reprint,a4paper,12pt,onecolumn,amsmath,amssymb,aps,prd,notitlepage,longbibliography,floatfix,nofootinbib]{revtex4-1}
\usepackage{amsmath}
\usepackage[utf8]{inputenc}
\usepackage{latexsym}
\usepackage{amsfonts}
\usepackage{graphicx,color}
\usepackage{mathrsfs}
\usepackage{epstopdf}
\usepackage{subfigure}

\usepackage{booktabs}
\usepackage{multirow}
\usepackage{soul}
\usepackage{microtype}
\usepackage{xcolor}     

\begin{document}
\title{Constraints on the Primordial Curvature Power Spectrum by Pulsar Timing Array: A Polynomial Parameterization Approach}

\author{Qin Fei}
\email{feiqin@hbpu.edu.cn }
\affiliation{School of Mathematics and Physics, Hubei Polytechnic University, Huangshi 435003, China}

\begin{abstract}
The recent stochastic signal observed jointly by NANOGrav, PPTA, EPTA, and CPTA can be accounted for by scalar-induced gravitational waves (SIGWs).   The source of the SIGWs is from the primordial curvature perturbations, and the main contribution to the SIGWs is from  
the peak of the primordial curvature power spectrum. To effectively model this peak, we apply the Taylor expansion to parameterize it.  With the Taylor expansion parameterization, we apply Bayesian methods to constrain the primordial curvature power spectrum based on the NANOGrav 15-year data set.  The constraint on the primordial curvature power spectrum possesses a degree of generality, as the Taylor expansion can effectively approximate a wide range of function profiles.
\end{abstract}

\maketitle
\section{Introduction}
The detection of gravitational waves (GWs) from compact binary mergers~\cite{LIGOScientific:2018mvr,LIGOScientific:2020ibl,LIGOScientific:2021djp} by LIGO-Virgo-KAGRA has provided valuable insights into the properties of the population of GW sources~\cite{LIGOScientific:2018jsj,Chen:2018rzo,Chen:2019irf,LIGOScientific:2020kqk,Chen:2021nxo,KAGRA:2021duu,Chen:2022fda,Liu:2022iuf,Sasaki:2021iuc, Kimura:2021sqz, Wang:2022nml,  Zheng:2022wqo,You:2023ouk}. 
The next crucial milestone in GW detection is the observation of stochastic GW backgrounds that would provide valuable information about astrophysical and cosmological processes, such as the early universe, dark matter~\cite{Wei:2013et,Wei:2013rea,Du:2023zsz}, and modified gravity~\cite{Chen:2014qsa,Wu:2015naa,Huang:2015yva,Zhu:2015lry,Gong:2017kim}.
Recently, the North American Nanohertz Observatory for Gravitational Waves (NANOGrav) \cite{NANOGrav:2023gor,NANOGrav:2023hde}, Parkes Pulsar Timing Array (PPTA) \cite{Zic:2023gta,Reardon:2023gzh}, European Pulsar Timing Array (EPTA)  together with  Indian Pulsar Timing Array (InPTA)   \cite{Antoniadis:2023lym,Antoniadis:2023ott}, and Chinese Pulsar Timing Array (CPTA) \cite{Xu:2023wog}, have jointly detected a common-spectrum process exhibiting the Hellings-Downs angular correlation feature inherent in gravitational waves (GWs). Characterized by a fiducial $f^{-2/3}$ characteristic-strain spectrum,
this signal is attributed to an ensemble of binary supermassive black hole inspirals, with a strain amplitude of approximately $\sim 10^{-15}$ at a reference frequency of $1\,\rm{yr}^{-1}$ \cite{NANOGrav:2023gor,Reardon:2023gzh,Antoniadis:2023ott,Xu:2023wog}. Besides the supermassive black hole binary (SMBHB)
scenario, other hypotheses are also proposed   ~\cite{NANOGrav:2023hvm,Antoniadis:2023xlr,Yi:2023mbm,Franciolini:2023pbf,Liu:2023ymk,Vagnozzi:2023lwo,Cai:2023dls,Bi:2023tib,Wu:2023hsa,Franciolini:2023wjm,You:2023rmn,Jin:2023wri,Liu:2023pau,An:2023jxf,Zhang:2023nrs,Das:2023nmm,Balaji:2023ehk,Du:2023qvj,Oikonomou:2023qfz,Yi:2023npi,Yi:2023tdk,Chen:2023zkb,Liu:2023hpw,Guo:2023hyp}. 
Among these hypotheses, scalar-induced gravitational waves (SIGWs) are a strong candidate, offering a more robust explanation than the supermassive black hole binary (SMBHB) model by Bayesian analysis \cite{NANOGrav:2023hvm, Yi:2023mbm}.   This paper focuses on exploring the observed PTA signal explained by SIGWs. For further physical processes contributing to the Pulsar Timing Array (PTA) band, please refer to~\cite{Zhu:2018lif,Li:2019vlb,Chen:2021wdo,Wu:2021kmd,Chen:2021ncc,Chen:2022azo,PPTA:2022eul,IPTA:2023ero,Wu:2023pbt,Wu:2023dnp,IPTA:2023ero,InternationalPulsarTimingArray:2023mzf,Wu:2023rib,Bi:2023ewq,Chen:2023uiz,Chen:2023bms,Wang:2023len}.

The scalar-induced gravitational waves are induced by the large scalar perturbations seeded from primordial curvature perturbations generated during the inflationary epoch \cite{Ananda:2006af,Baumann:2007zm,Kohri:2018awv}. Accompanying the formation of SIGWs, the large scalar perturbations can produce primordial black holes (PBHs) 
 \cite{Hawking:1971ei,Carr:1974nx,Saito:2008jc,Alabidi:2012ex,Sasaki:2018dmp,Nakama:2016gzw,Di:2017ndc,Cheng:2018yyr,Cai:2019amo,Cai:2018dig,Cai:2019elf,Cai:2019bmk,Wu:2020drm,Cai:2020fnq,Pi:2020otn,Domenech:2020kqm,Liu:2018ess,Liu:2019rnx,Liu:2020cds,Liu:2021jnw,Liu:2022iuf,Chen:2019xse,Yuan:2019fwv,Yuan:2019wwo,Yuan:2019udt,Carr:2020gox,Liu:2020vsy,Liu:2020bag,Yi:2022ymw,Papanikolaou:2021uhe,Papanikolaou:2022hkg,Chakraborty:2022mwu,Liu:2022cuj,Chen:2018czv,Chen:2018rzo,Chen:2019irf,Chen:2021nxo,Chen:2022fda,Liu:2022wtq,Zheng:2022wqo,Chen:2022qvg,Meng:2022low}. To produce detectable SIGWs, the power spectrum amplitude of primordial curvature perturbations, denoted as $\mathcal{A}_{\zeta}$,  must be on the order of $\mathcal{A}_{\zeta} \sim \mathcal{O}(0.01)$, while the constraints 
 from cosmic microwave background (CMB) anisotropy observations at large scales are $\mathcal{A}_{\zeta} = 2.1\times 10^{-9}$ \cite{Planck:2018jri}.  Therefore, the significant SIGWs require the power spectrum amplitude of primordial curvature perturbations seven orders of magnitude larger than the constraints at large scales.  Generating such a large enhancement in the primordial curvature power spectrum is hard to achieve in traditional slow-roll inflation models;  instead, ultra-slow-roll inflationary scenarios may be able to achieve this purpose
 \cite{Martin:2012pe,Motohashi:2014ppa,Yi:2017mxs,Yi:2016jqr,Fei:2017fub, Garcia-Bellido:2017mdw,Germani:2017bcs,Motohashi:2017kbs,Ezquiaga:2017fvi,Gong:2017qlj,Yi:2018dhl,Yi:2018gse,Ballesteros:2018wlw,Dalianis:2018frf,Bezrukov:2017dyv,Kannike:2017bxn,Fei:2020jab,Lin:2020goi,Lin:2021vwc,Yi:2021xhw,Gao:2020tsa,Gao:2019sbz,Gao:2021vxb,Yi:2020kmq,Yi:2020cut,Yi:2021lxc,Yi:2022anu,Zhang:2020uek,Pi:2017gih,Kamenshchik:2018sig,Fu:2019ttf,Fu:2019vqc,Dalianis:2019vit,Gundhi:2020zvb,Cheong:2019vzl,Zhang:2021rqs,Zhang:2021vak,Kawai:2021edk,Cai:2021wzd,Chen:2021nio,Karam:2022nym,Ashoorioon:2019xqc,Garcia-Saenz:2023zue}. 

The energy density of SIGWs is mainly determined by the peak structure of the primordial curvature power spectrum. In this study, we adopt a polynomial representation to characterize the peak profile of the primordial curvature power spectrum. This polynomial form can be understood as a Taylor expansion capable of effectively describing various peak shapes. By employing this polynomial parameterization, we can derive the most data-favored peak profile of the primordial curvature power spectrum by the Bayes analysis. 
The organization of this paper is as follows. 
Section \ref{sec:sigw} gives the energy density of the SIGWs. Section \ref{sec:result} overviews the parameterization of the primordial curvature power spectrum and presents the constraints on it. Finally, the conclusions are drawn in Section \ref{sec:con}.

\section{The scalar-induced gravitational waves}\label{sec:sigw}
If the scalar perturbation resulting from the primordial curvature perturbations created during inflation reaches a notable level, the secondary order of linear scalar perturbations may act as a source to induce GWs.  In this section, we explore a detailed analysis of the energy density of the SIGWs during the radiation dominated.
 The perturbed metric within the Newtonian gauge can be  expressed as follows:
\begin{equation}
{\rm d} s^2=a^2(\eta)\left[-(1+2\Phi){\rm d} \eta^2+\left\{(1-2\Phi)\delta_{ij}+\frac{1}{2}h_{ij}\right\}{\rm d} x^i {\rm d} x^j\right].
\end{equation}
Here, ${\Phi}$ is the Bardeen potential, and $h_{ij}$ represents the tensor perturbation, satisfying the transverse-traceless gauge condition $\partial^{i}h_{ij}=h^{i}_{i}=0$. In the Fourier space, the tensor perturbation is denoted by:
\begin{equation}
h_{ij}\left(\boldsymbol{x},\eta\right)=\int\frac{{\rm d} ^3\boldsymbol{k}}{\left(2\pi\right)^{3/2}}\text{e}^{{\rm i} \boldsymbol{k}\cdot\boldsymbol{x}}\left[h^{+}_{\boldsymbol{k}}\left(\eta\right)e^{+}_{ij}\left(\boldsymbol{k}\right)
+h^{\times}_{\boldsymbol{k}}\left(\eta\right)e^{\times}_{ij}\left(\boldsymbol{k}\right)\right],
\end{equation}
where ${e^{+}_{ij}\left(\boldsymbol{k}\right)}$ and ${e^{\times}_{ij}\left(\boldsymbol{k}\right)}$ are the plus and cross-polarization tensors,
\begin{equation}
\begin{split}
 &e^{+}_{ij}\left(\boldsymbol{k}\right)=\frac{1}{\sqrt{2}}\left[e_i\left(\boldsymbol{k}\right)e_j\left(\boldsymbol{k}\right)
 -\overline{e}_i\left(\boldsymbol{k}\right)\overline{e}_j\left(\boldsymbol{k}\right)\right],\\& 
 e^{\times}_{ij}\left(\boldsymbol{k}\right)=\frac{1}{\sqrt{2}}\left[e_i\left(\boldsymbol{k}\right)
\overline{e}_j\left(\boldsymbol{k}\right)+\overline{e}_i\left(\boldsymbol{k}\right)e_j\left(\boldsymbol{k}\right)\right].
 \end{split}
\end{equation}
Here ${e_{i}\left(\boldsymbol{k}\right)}$ and  ${\overline{e}_{i}\left(\boldsymbol{k}\right)}$ are the normalized basis vectors orthogonal to the wave vector $\boldsymbol{k}$, with ${\boldsymbol{e}\cdot\boldsymbol{\overline{e}} =\boldsymbol{e}\cdot\boldsymbol{k}=\boldsymbol{\overline{e}}\cdot\boldsymbol{k}=0}$.

In Fourier space and neglecting anisotropic stress,  the equation of motion for tensor perturbations with either polarization can be expressed as: 
\begin{equation}\label{E4}
h^{''}_{\boldsymbol{k}}\left(\eta\right)+2\mathcal{H}h^{'}_{\boldsymbol{k}}\left(\eta\right)+k^2h^{}_{\boldsymbol{k}}\left(\eta\right)=4S_{\boldsymbol{k}}\left(\eta\right),
\end{equation}
where $\eta$ represents the conformal time, ${\rm d}  \eta = {\rm d} t/a$, ${\mathcal{H}=a'/a}$ is the conformal Hubble parameter.  The source term from the linear scalar perturbations is denoted by:
\begin{equation}
\label{hksource}
 S_{\boldsymbol{k}}=\int \frac{{\rm d} ^3\boldsymbol{\tilde{k}}}{(2\pi)^{3/2}}e_{ij}(\boldsymbol{k})\tilde{k}^i\tilde{k}^j
 \left[2\Phi_{\tilde{\boldsymbol{k}}}\Phi_{\boldsymbol{k}-\tilde{\boldsymbol{k}}} \phantom{\frac{1}{2}}+ 
 \frac{1}{\mathcal{H}^2} \left(\Phi'_{\tilde{\boldsymbol{k}}}+\mathcal{H}\Phi_{\tilde{\boldsymbol{k}}}\right)
 \left(\Phi'_{\boldsymbol{k}-\tilde{\boldsymbol{k}}}+\mathcal{H}\Phi_{\boldsymbol{k}-\tilde{\boldsymbol{k}}}\right)\right],
 \end{equation}
where $\Phi_{\boldsymbol{k}}$  denotes the  Bardeen potential in the Fourier space, which can be related to the  primordial curvature perturbations by 
\begin{equation}
 \Phi_{\boldsymbol{k}}=\frac{3+3w}{5+3w}T(k\eta) \zeta_{\boldsymbol{k}},
 \end{equation}
and  the transfer function $T(x)$ can be expressed as 
\begin{equation}\label{transfer}
   T(x) =3\left[\frac{\sin(x/\sqrt{3})-(x/\sqrt{3}) \cos(x/\sqrt{3})}{(x/\sqrt{3})^3}\right]. 
\end{equation}
By using the Green's function method, we can obtain the solution for $h_{\boldsymbol{k}}\left(\eta\right)$, 
\begin{equation}\label{E7}
h_{\boldsymbol{k}}\left(\eta\right)=\frac{4}{a\left(\eta\right)}\int^{\eta}{\rm d} \overline{\eta}G_{\boldsymbol{k}}\left(\eta,\overline{\eta}\right)
a\left(\overline{\eta}\right)
S_{\boldsymbol{k}}\left(\overline{\eta}\right),
\end{equation}
with  the Green function 
 \begin{equation}
    G_{\boldsymbol{k}}\left(\eta,\overline{\eta}\right)=\frac{\sin\left[k\left(\eta-\overline{\eta}\right)\right]}{k}.
\end{equation}
The definition of the power spectrum of SIGWs, $\mathcal{P}_h$, is
\begin{equation}\label{Ep}
\langle h_{\boldsymbol{k}}\left(\eta\right)h_{\boldsymbol{p}}\left(\eta\right)\rangle
=\delta^3\left(\boldsymbol{k}+\boldsymbol{p}\right) \frac{2\pi^2}{k^3}\mathcal{P}_h\left(k,\eta\right).
\end{equation}
Combining  the above relations, we can obtain 
\cite{Kohri:2018awv,Lu:2019sti,Baumann:2007zm,Inomata:2016rbd,Espinosa:2018eve,Ananda:2006af}
 \begin{equation}\label{Eph2}
 \begin{split}
 \mathcal{P}_h(k,\eta)= 
 4\int_{0}^{\infty}{\rm d} v\int_{|1-v|}^{1+v}{\rm d} u \left[\frac{4v^2-(1-u^2+v^2)^2}{4uv}\right]^2   I_{RD}^2(u,v,x)\mathcal{P}_{\zeta}(kv)\mathcal{P}_{\zeta}(ku).
 \end{split}
 \end{equation}
Here,  the new variables are ${u=|\boldsymbol{k}-\tilde{\boldsymbol{k}}|/k,  v=\tilde{|\boldsymbol{k}|}/k}$, $x= k \eta$, and  $\mathcal{P}_\zeta$ is the  power spectrum of the primordial curvature. Furthermore, we have 
 \begin{equation}
 	\label{irdeq1}
 	\begin{split}
 		I_{\text{RD}}(u, v, x)=&\int_0^x {\rm d} y\, y \sin(x-y)\{3T(uy)T(vy)\\
 		&+y[T(vy)u T'(uy)+v T'(vy) T(uy)]\\
 		&+y^2 u v T'(uy) T'(vy)\}.
 	\end{split}
 \end{equation}
By using  the definition of the  energy density parameter of GWs 
 \begin{equation}
 	\label{density}
 	\Omega_{\mathrm{GW}}(k,\eta)=\frac{1}{24}\left(\frac{k}{aH}\right)^2\overline{\mathcal{P}_h(k,\eta)}.
 \end{equation}
and equations \eqref{Eph2}, we can obtain \cite{Espinosa:2018eve,Lu:2019sti}
 \begin{equation}
 	\label{SIGWs:gwres1}
 	\begin{split}
 		\Omega_{\mathrm{GW}}(k,\eta)=&\frac{1}{6}\left(\frac{k}{aH}\right)^2\int_{0}^{\infty}{\rm d} v\int_{|1-v|}^{1+v}{\rm d} u  
 		  \left[\frac{4v^2-(1-u^2+v^2)^2}{4uv}\right]^2 
 		 \overline{I_{\text{RD}}^{2}(u, v, x)} \mathcal{P}_{\zeta}(kv)\mathcal{P}_{\zeta}(ku),
 	\end{split}
 \end{equation}
where $\overline{I_{\text{RD}}^{2}}$ is the oscillation time average of $I_{\text{RD}}^{2}$. After the horizon reentry and at late time $x\rightarrow\infty$, $\overline{I_{\text{RD}}^{2}}$ becomes \cite{Kohri:2018awv}
\begin{equation}\label{IRD}
\begin{split}
 \overline{I_{\mathrm{RD}}^2(v,u,x\rightarrow \infty)} & = \frac{1}{2x^2} \left( \frac{3(u^2+v^2-3)}{4 u^3 v^3 } \right)^2  \times \left\{\pi^2 (u^2+v^2-3)^2 \Theta \left( v+u-\sqrt{3}\right) \right. \\& \left. -\left(4uv-(u^2+v^2-3) \ln \left| \frac{3-(u+v)^2}{3-(u-v)^2} \right| \right)^2 \right\}.
\end{split}
\end{equation}
The evolution of GW energy density is the same as that of radiation. By using this characteristic, we can  get the present energy density of GWs:
\begin{equation}\label{d}
\Omega_{\mathrm{GW}}(k,\eta_0)= \Omega_{r,0} \Omega_{\mathrm{GW}}(k,\eta),
\end{equation}
where $\Omega_{r,0}$ is the energy density parameter of radiation at present.

\section{Methodology and Results}\label{sec:result}
The contribution to the energy density of SIGWs is mainly from the peak region of the primordial curvature power spectrum.  In this paper, we focus on the peak region of the primordial curvature power spectrum and  parameterize it as 
\begin{equation}\label{pr:para1}
    \log_{10}\mathcal{P}_\zeta(k) = \sum_{i=0}^{n} (\log_{10}A_i)\cdot x^i, \quad -1<x<1,
\end{equation}
where $x = \log_{10}(k/k_p)$, and $k_p$ is the peak scale of primordial curvature power spectrum. 
The peak can be well described by the polynomial as long as the polynomial term is large enough. The Taylor expansion converges at the condition  $|x|<1$ . Because larger polynomial terms will cost larger computing resources, for the sake of simplification, in this paper, we only consider $n=3$. 
To ensure that the peak of the power spectrum is located at the scale $k_p$, the derivatives of power spectrum must satisfy ${\rm d} \mathcal{P}_\zeta/{\rm d}x = 0$ and ${\rm d}^2 \mathcal{P}_\zeta/{\rm d} x^2 <0$ at   $x=0$. Furthermore, to ensure that the peak at scale $k_p$ is the only peak, the first derivative should satisfy ${\rm d} \mathcal{P}_\zeta/{\rm d}x > 0$ for $x<0$  and ${\rm d} \mathcal{P}_\zeta/{\rm d}x < 0$ for $x>0$. For the parameterization considered in this paper, these conditions are equivalent to
\begin{equation} 
\log_{10} A_1 = 0, \quad \log_{10} A_2 <-1.5 ~ |\log_{10} A_3|.
\end{equation}

\begin{figure}[htbp!] 
\centering
\includegraphics[width=0.8\textwidth]{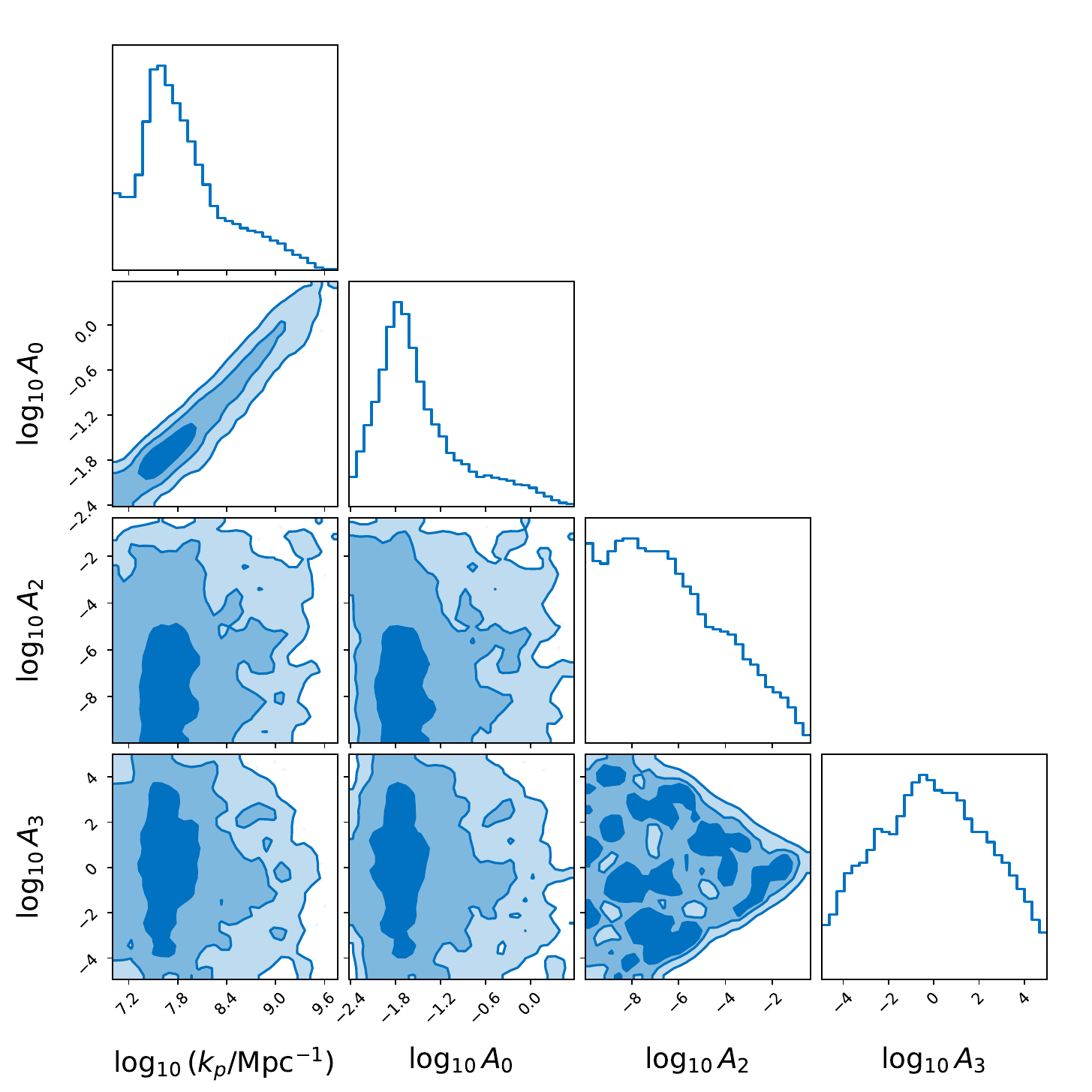}
\caption{The posterior of the parameters in parameterization \eqref{pr:para} using the  NANOGrav 15-yrs data set with Bayesian methods. }
\label{fig:free_Pk}
\end{figure}

\begin{figure}[htbp!] 
\centering
\includegraphics[width=0.8\textwidth]{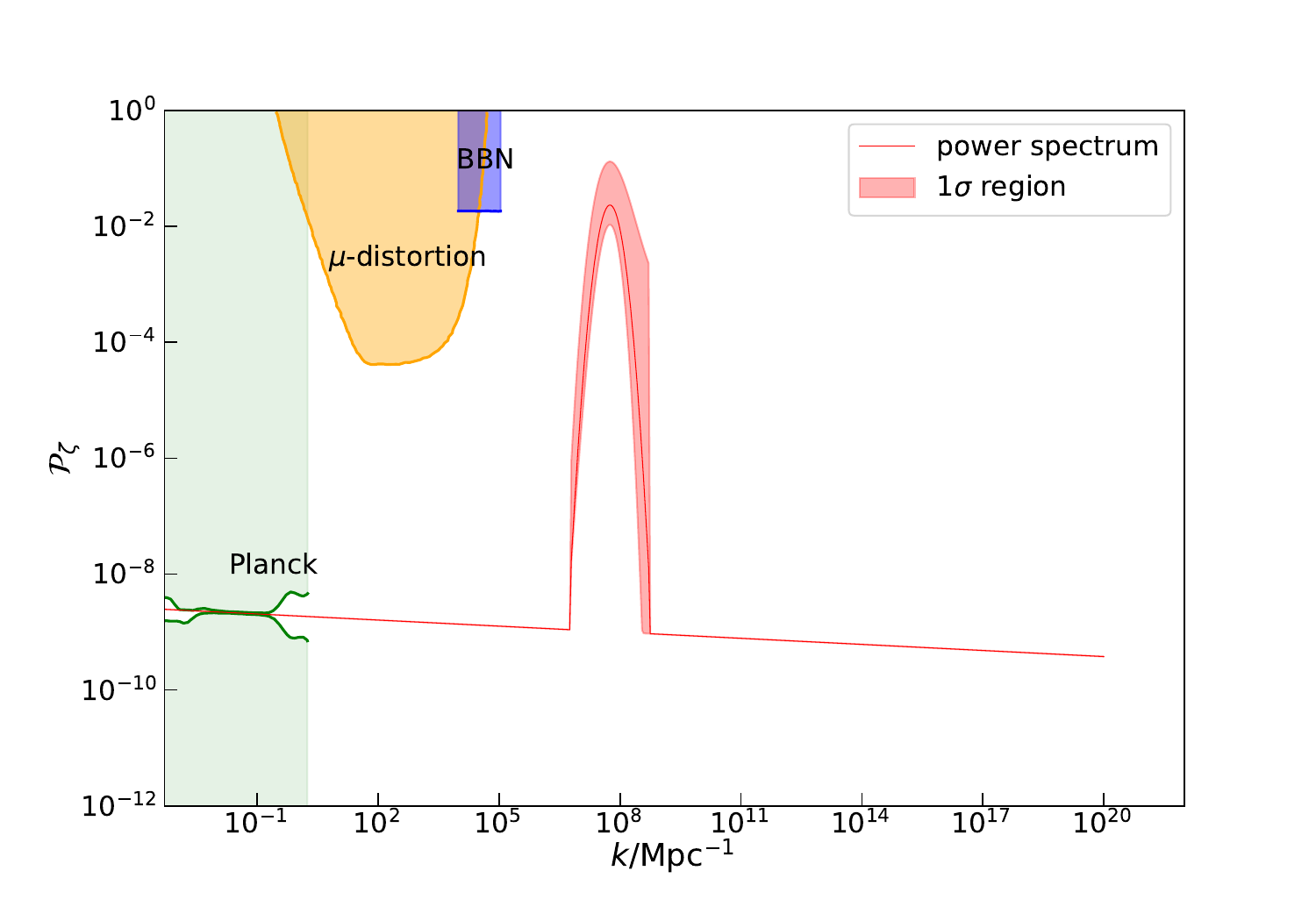}
\caption{The primordial power spectrum of the parameterization \eqref{pr:para}  by choosing the 1-$\sigma$ value listed in Table. \ref{Tab:prior}. 
The green, orange, and blue regions are the constraints from the Planck data \cite{Planck:2018jri}, $\mu$-distortion of CMB \cite{Fixsen:1996nj}, and the effect on the ratio between neutron
and proton during the Big Bang nucleosynthesis (BBN) \cite{Inomata:2016uip}, respectively.}
\label{fig:pr}
\end{figure}

\begin{figure}[htbp!] 
\centering
\includegraphics[width=0.8\textwidth]{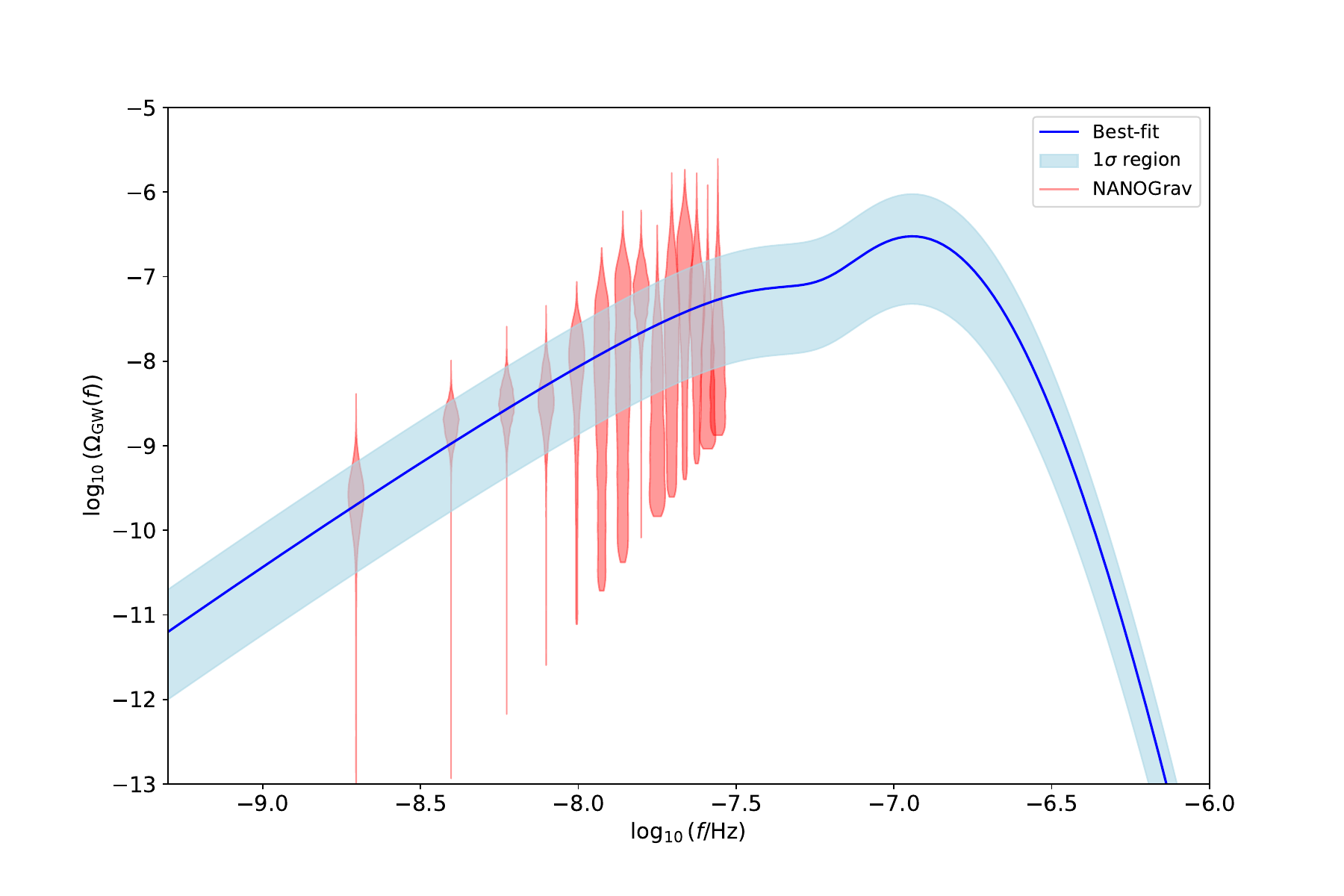}
\caption{The corresponding energy density of the SIGWs denoted by the blue line.  The red violins are the NANOGrav 15 yrs data set.  }
\label{fig:gw}
\end{figure}

For other scales, we use the near-scale-invariance power-law form to parameterize the primordial curvature power spectrum, 
\begin{equation}\label{pr:para2}
\mathcal{P}_\zeta(k) = A_* \left(\frac{k}{k_*}\right)^{n_s-1}, \quad x\leq-1 ~~\mathrm{or}~~ x\geq 1,
\end{equation}
where $k_*=0.05\mathrm{Mpc}^{-1}$,  $A_* = 2.1\times 10^{-9}$, and $n_s =0.965$ \cite{Planck:2018jri}.   Therefore, the parameterization is 
\begin{equation}\label{pr:para}
\log_{10}\mathcal{P}_\zeta(k) 
=\begin{cases}
\sum_{i=0}^{n} (\log_{10}A_i)\cdot x^i,& -1<x<1,\\
\log_{10} \left[A_* \left(\frac{k}{k_*}\right)^{n_s-1}\right],&x\leq-1 ~~\mathrm{or}~~ x\geq 1.
\end{cases}
\end{equation}
There are steps present at $x=-1$ and $x=1$ in this parameterization. These steps in the power spectrum of the primordial curvature power spectrum can be found in various inflationary mechanisms, such as parametric amplification \cite{Cai:2019bmk} and sound speed resonance \cite{Cai:2018tuh}. Hence, the inclusion of steps in the parameterization \eqref{pr:para} is considered acceptable.  
For the parameterization given by \eqref{pr:para}, when $|x|<1$, if the spectrum obtained from parameterization \eqref{pr:para1} is smaller than the spectrum obtained from parameterization \eqref{pr:para2}, we use the latter.
In the following, we use the parameterization described by Eq. \eqref{pr:para} and apply Bayesian methods to the NANOGrav 15-yrs data to obtain the constraint on the amplitude of the primordial curvature power spectrum.  Our analysis utilizes data from 14 frequency bins in the NANOGrav 15-year data set \cite{NANOGrav:2023gor, NANOGrav:2023hvm} to infer the posterior distributions of the parameters in parameterization \eqref{pr:para}. The computations were performed using the \texttt{Bilby} code \cite{Ashton:2018jfp}, implementing the \texttt{dynesty} algorithm for nested sampling  \cite{NestedSampling}.  
We formed the log-likelihood function by evaluating the energy density of SIGWs at the 14 specific frequency intervals. Afterward, we computed the sum of the logarithms of probability density functions obtained from 14 independent kernel density estimates \cite{Moore:2021ibq,Lamb:2023jls,Liu:2023ymk,Wu:2023hsa,Jin:2023wri,Liu:2023pau}. Consequently, we can represent the likelihood function as  
\begin{equation}
\ln \mathcal{L}(\Lambda)=\sum_{i=1}^{14}\ln \mathcal{L}_i\left(\Omega_{\mathrm{GW}}\left(f_i, \Lambda \right)\right),
\end{equation}
where $\Lambda$ denotes all the  parameters in the parameterization given by Eq.  \eqref{pr:para}, and the prior for the parameters are listed in Table \ref{Tab:prior}. The function $\mathcal{U}$ denotes the uniform distribution. 
In this study, we incorporate constraints stemming from baryon acoustic oscillation and cosmic microwave background (CMB) data~\cite{Planck:2018vyg}. Specifically, we impose a restriction on the integrated energy-density fraction, given by $\int_{k_{\min }}^{\infty} h^2 \Omega_{\mathrm{GW}, 0}(k), d\ln k\lesssim 2.9 \times 10^{-7}$~\cite{Clarke:2020bil}, where $h = H_0 / ( 100 \mathrm{km}, \mathrm{s}^{-1}, \mathrm{Mpc}^{-1})=0.674$~\cite{Planck:2018vyg} is the dimensionless Hubble constant.
The posterior distributions of the parameters in the parameterization \eqref{pr:para}  are displayed in Figure \ref{fig:free_Pk}, and the maximum posterior values, 1-$\sigma$ credible interval bounds of posteriors are listed in Table \ref{Tab:prior}. The maximum posterior of the peak scale is   $ k_p = 5.6 \times 10^{7} \mathrm{Mpc}^{-1}$, which is consistent with the Gauss model given in the NANOGrav paper \cite{NANOGrav:2023hvm}.  
\begin{table}[htbp]
\centering
\begin{tabular}{l|llllll}
\hline
\hline
Parameters & $\log_{10} (k_p/\mathrm{Mpc}^{-1})$ & $\log_{10} A_0$&  $\log_{10} A_2$& $\log_{10} A_3$ \\
\hline
prior & $\mathcal{U}(7,10)$ & $\mathcal{U}(-4,1)$&  $\mathcal{U}(-10,2)$& $\mathcal{U}(-5,5)$ \\
\hline
posterior & $7.75^{+0.68}_{-0.34}$ &  $-1.63^{+0.75}_{-0.34}$&  $-6.74^{+2.98}_{-2.17}$& $-0.07^{+2.63}_{-2.62}$\\
\hline 
\hline 
\end{tabular}
\caption{The priors, maximum posterior values, 1-$\sigma$ credible interval bounds of posteriors for the parameters in the primordial curvature power spectrum parameterization \eqref{pr:para} by using NANOGrav 15-yrs data set and Bayesian methods.}
\label{Tab:prior}
\end{table}
By choosing the values in the 1-$\sigma$ credible interval bounds of the parameters listed in Table \ref{Tab:prior}, we calculate the primordial curvature power spectrum shown in Figure \ref{fig:pr} and denoted by the red line;  the corresponding SIGW are displayed in the   Figure \ref{fig:gw} and denoted by the blue line.  Figure \ref{fig:gw} shows that the energy density of the SIGWs is consistent with the NANOGrav 15-yrs data set. 
\section{Conclusion}\label{sec:con}
The stochastic signal detected by the NANOGrav, PPTA, EPTA, and CPTA collaborations can be explained by SIGWs, where the source of the GWs is from the primordial curvature perturbations. 
The main contribution to the energy density of the SIGWs is from the peak of the primordial curvature power spectrum. In this paper, we focus on the peak region and parameterize the primordial curvature power spectrum near the peak by the Taylor expansion with four terms to characterize the peak behavior. The posterior of the parameters are:  $\log_{10} (k_p/\mathrm{Mpc}^{-1})=7.75^{+0.68}_{-0.34}$, $\log_{10} A_0=-1.63^{+0.75}_{-0.34}$,  $\log_{10} A_2=-6.74^{+2.98}_{-2.17}$, $\log_{10} A_3=-0.07^{+2.63}_{-2.62}$.
Because the Taylor expansion can describe most function profiles, our constraint on the primordial curvature power spectrum possesses a degree of generality.


\section*{Acknowledgments}
This research was supported in part by the National Natural Science Foundation of China under Grant No. 12305060, and the Talent-Introduction Program of Hubei Polytechnic University under Grant No.19xjk25R.



%

\end{document}